\documentstyle[aps,prl,multicol]{revtex}

\draft

\begin{document}

\title{On the equivalence between the Barkhausen effect and\\
directed Abelian sandpile models}
\author{Alexei V\'azquez and Oscar Sotolongo-Costa}
\address{Department of Theoretical Physics, Faculty of
Physics, Havana University, Havana 10400, Cuba}

\maketitle

\begin{abstract}

The existence of self-organized criticality in the Barkhausen
effect and its analogy with sandpile models is investigated. It
is demonstrated that a model recently introduced to describe the
dynamics of a domain wall [Cizeau {\em et al}, Phys. Rev. Lett.
{\bf 79}, 4669 (1997)] belongs to the universality class of
undirected Abelian sandpile models. In this way it is shown that
the Barhausen effect can be taken as an experimental observation
of self-organized critical phenomena.

\end{abstract}

\pacs{75.60.Ej, 05.40.+j, 64.60.Lx, 68.35.Rh}

\begin{multicols}{2}

One of the fundamental tasks of the theory of critical phenomena
is to determine the different universality classes. If different
systems or phenomena cannot be grouped into a reduced group of
universality classes then the central idea of critical
phenomena, the existence of universal behavior, will be of no
relevance. This task has been carry out succesfully in ordinary
critical phenomena. However, the precise identification of the
universality classes in a non-equilibrium critical phenomena
like self-organized criticality (SOC) is still unresolved.

SOC was introduced to explain the critical behavior of a vast
class of driven dissipative system which evolve into a critical
state \cite{bak}. In its early state it was believed that such a
critical state is insensitive to changes in control parameters
and no fine-tuning is needed. More recent interpretations of
this phenomena have shown that criticality in SOC systems is
obtained after some control parameters, for instance the driving
and dissipation rates, are fine-tuned to zero \cite{vespignani}.

The Barkhausen effect has been taken as an experimental
observation of SOC behavior. Based on phenomenological analogies
between the Barkhausen effect and sandpile models, such as the
existence of power law distributions of avalanche size and
duration, some authors have claim that the Barkhausen effect
exhibits SOC behavior \cite{geoffroy,cote}. However, this
conclusion has been criticized by other researches which pointed
out that the observation of power law distributions is not
necessarely an evidence of SOC behavior
\cite{obrien,spasojevic}. There are alternative approaches, like
the random field Ising model \cite{sethna,perkovic,dahmen},
where the power law distributions are a consequence of the
scaling properties of disorder. The determination of the
universality classes in this case thus become more difficult
because it is still not clear if the Barkhausen effect exhibits
or not SOC behavior.

Some light in this controversy has been given by Cizeau,
Zapperi, Durin and Stanley (CZDE) \cite{cizeau}. They introduced
an equation of motion for a single domain wall where dipolar
interactions, demagnetization effects and quenched disorder are
considered, containing some previous works
\cite{alessandro,urbach} as limiting cases. CZDE observed that
their model has certain analogies with sandpile models. More
precisely the critical state is obtained when the magnetic field
rate (the driving rate in sandpile models) and the
demagnetization factor (the dissipation rate in sandpile models)
are fine-tunned to zero. However their analysis, in relation to
this analogy, was limited to this phenomenological observation
based on numerical simulations and mean field analysis. The
great importance of this conjecture is that if one could map the
Barkhausen effect into certain class of sandpile models then one
could be sure about the existence of SOC in this phenomena. This
is precisely the scope of this work.

We investigate the CZDE model in the case of strong
magnetization, where dipolar interactions are relevant
\cite{note}. First we show that when the magnetization field
increases at constant rate the domain wall is never pinned but
moves in average at constant velocity. Then we obtain exact
expressions for the average interface velocity and
susceptibility. The scaling exponents are obtained using
perturvative analysis, some of them but not all results
identical to those obtained by CZDE for the case of constant
magnetic field. Further RG calculations reveals that
perturbation theory is exact up to ${\cal O}(r^{-1})$, where $r$
is proportional to the square of the saturation magnetization.
From the comparison of the scaling exponents with those observed
in sandpile models we conclude that the CZDE model with magnetic
field increasing at constant rate belongs to the universality
class of directed Abelian sandpile models (DASM) \cite{dhar}.

To start our analysis let us introduce the CZDE model. The
domain wall is modeled by a $d$-dimensional interface, dividing
two regions of opposite magnetization, moving in a $d+1$
environment described by its position $h(\vec{x},t)$.
Considering the contribution of magnetostatic, ferromagnetic and
magneto-chrystalline interactions one obtains the following
equation of motion \cite{cizeau}
\begin{eqnarray} 
\lambda \frac{\partial}{\partial t} h(\vec{x},t)
=\Gamma\nabla^2h(\vec{x},t)+2\mu_0M_sH
\nonumber\\
+\eta[\vec{x},h(\vec{x},t)]-
4\mu_0{\cal N}M_s^2\int\frac{d^dx^\prime}{L^d}h(\vec{x}^\prime,t)
\nonumber\\
+\int d^dx^\prime K(\vec{x}-\vec{x}^\prime)[h(\vec{x}^\prime,t)
-h(\vec{x},t)],
\label{eq:1} 
\end{eqnarray}
where $\lambda$ is a viscosity coefficient, $\Gamma$ is the
surface tension of the wall, $H$ is the magnetic field
intensity, $L$ is the linear size of the system, and $M_s$ is
the saturation magnetization per unit volume. Long-range
demagnetization effects are described by the fourth term in the
right hand side, where ${\cal N}$ is the demagnetization factor.
Dipolar interactions are characterized by the fifth term, where
the kernel $K(\vec{x})$ is anisotropic and has Fourier transform
\begin{equation}
\tilde{K}(\vec{k})=\frac{\mu_0M_s^2}{4\pi^2}|\vec{k}|\cos^2\theta,
\label{eq:2}
\end{equation}
where $\theta$ is the angle between $\vec{k}$ and the
magnetization. $\eta(\vec{x},h)$ is a Gaussian uncorrelated
noise due to lattice defects or other factors, with zero mean
and noise correlator
\begin{equation}
\langle\eta(\vec{x},h)\eta(\vec{x}^\prime,h^\prime)\rangle =
\delta^d(\vec{x}-\vec{x}^\prime)\Delta(h-h^{\prime}), 
\label{eq:3} 
\end{equation}
where $\Delta(h)$ is a monotonically decreasing even function.

If dipolar interactions and demagnetization effects are
neglected and $H$ is constant then eq. (\ref{eq:1}) is reduced
to the Edwards-Wilkinson equation with quenched noise. This
limiting case has been extensively studied in the literature
\cite{fisher,review}. A depinning transition takes place at
certain critical field $H_c$, determined by the disorder. For
$H<H_c$ the interface is pinned after certain finite time while
for $H>H_c$ it moves with finite average velocity. The upper
critical dimension is $d_c=4$. This features remains if dipolar
interactions are considered but the upper critical dimension is
reduced to $d_c=2$ \cite{cizeau}.

When the demagnetization field is included and the magnetic
field increases at rate $c$ then the interface is never pinned
by impurities, but always moves with a finite average velocity
$v$. A perturbative solution of eq. (\ref{eq:1}) can thus be
found expanding $h(\vec{x},t)$ around the flat co-moving
interface $vt$. Taking $h(\vec{x},t)=vt+y(\vec{x},t)$, with
$\langle y(\vec{x},t)\rangle=0$, we obtain the following
equation for $y(\vec{x},t)$
\begin{eqnarray} 
\lambda \frac{\partial}{\partial t}y(\vec{x},t)
=\Gamma\nabla^2y(\vec{x},t)+2\mu_0M_s(c-2{\cal N}M_sv)t 
\nonumber\\
-\lambda v+\eta[\vec{x},vt+y(\vec{x},t)]-
4\mu_0{\cal N}M_s^2\int\frac{d^dx^\prime}{L^d}y(\vec{x}^\prime,t) 
\nonumber\\
+\int d^dx^\prime K(\vec{x}-\vec{x}^\prime)[y(\vec{x}^\prime,t)
-y\vec{x},t)],
\label{eq:4} 
\end{eqnarray}

The average velocity is obtained using the constraint $\langle
y(\vec{x},t)\rangle=0$. For this purpose is better to work with
the equation for the Fourier transform of $h(\vec{x},t)$,
$\tilde{h}(\vec{k}$,$\omega)$. The effective external field
$(c-2{\cal N}M_sv)t$ gives a singular term of the order of
$\omega^{-2}$. This singular term predominates in the low
frequency limit resulting, after imposing
$\langle\tilde{y}(\vec{k},\omega)\rangle=0$,
\begin{equation}
v=\frac{c}{2{\cal N}M_s}.
\label{eq:5}
\end{equation}
This result is valid to all orders of perturbation expansion
and, therefore, exact. In the MF theory by CZDE, which is
equivalent to the ABBM model, it is obtained that $v\sim c/{\cal
N}M_s$ in agreement with eq. (\ref{eq:5}). We have thus shown
that this result is exact and, therefore, valid beyond the MF
theory.

Another exact result can be obtained if one computes the
low-frequency and long-wavelength susceptibility. Adding a
source term $\varphi(\vec{x},t)$ to the right hand side of eq.
(\ref{eq:4}) and going to the Fourier space one obtains the
generalized response function
\begin{equation}
\tilde{G}(\vec{k},\omega)=\left\langle\frac{\tilde{h}(\vec{k},\omega)}
{\tilde{\varphi}(\vec{k},\omega)}\bigg|_{\tilde{\varphi}=0}\right\rangle=
\frac{1}{[\tilde{G}_0(\vec{k},\omega)]^{-1}-
\tilde{\Sigma}(\vec{k},\omega)},
\label{eq:6}
\end{equation}
where
\begin{equation}
[\tilde{G}_0(\vec{k},\omega)]^{-1}=\Gamma\vec{k}^2+
i\lambda\omega+4\mu_0{\cal N}M_s^2\hat{\delta}(\vec{k},L) 
+\tilde{K}(\vec{k}).
\label{eq:7}
\end{equation}
is the bare correlator and $\tilde{\Sigma}(k,\omega)$ is the
"self-energy". $\hat{\delta}(\vec{k},L)$ is the Fourier
transform of $L^{-d}$. In the thermodynamic limit
($L\rightarrow\infty$) $\hat{\delta}(\vec{k},L)\approx1$ for
$\vec{k}\rightarrow0$ and zero otherwise. Since
$\tilde{\Sigma}(0,0)=0$ and $\tilde{G}_0(0,0)^{-1}=4\mu_0{\cal
N}M_s^2$ it results that the low-frequency and long-wavelength
susceptibility (or simply the susceptibility) is given by
\begin{equation}
\chi=\tilde{G}(0,0)=\frac{1}{4\mu_0{\cal N}M_s^2}.
\label{eq:8}
\end{equation}
This result is also exact to all orders of perturbation
expansion.

In eq. (\ref{eq:8}) one cannot determine precisely which is the
control parameter of the model. The susceptibility may diverges
if both ${\cal N}$ or $M_s$ goes to zero. To answer this
question we look for a self-similar solution of eq.
(\ref{eq:4}). Performing the scale transformation $x\rightarrow
bx$, $L\rightarrow bL$, $t\rightarrow b^zt$ and $y\rightarrow
b^\zeta y$ where $z$ and $\zeta$ are the dynamic and roughness
exponent, respectively, and  taking into account eq.
(\ref{eq:5}), eq. (\ref{eq:4}) becomes
\begin{eqnarray}
\lambda b^{1-z} \frac{\partial}{\partial t}y(\vec{x},t)
=\Gamma b^{-1}\nabla^2y(\vec{x},t) -\lambda b^{1-\zeta} v
\nonumber\\
+b^{1-\zeta}\eta[b\vec{x},vb^zt+b^\zeta y(\vec{x},t)]
\nonumber\\
-4\mu_0{\cal N}M_s^2b\int\frac{d^dx^\prime}{L^d}y(\vec{x}^\prime,t) 
\nonumber\\
+\int d^dx^\prime K(\vec{x}-\vec{x}^\prime)[y(\vec{x}^\prime,t)
-y\vec{x},t)].
\label{eq:9} 
\end{eqnarray}
Notice that we cannot obtain scale invariance in the elastic and
dipolar intercation terms  simultaneously. Thus we have to
determine which of this terms gives the major contribution. A
magnitude which characterizes the ration between elastic and
dipolar interactions is
\begin{equation}
r=\frac{\mu_0M_s^2}{4\pi^2\Gamma\Lambda},
\label{eq:13}
\end{equation}
where $\Lambda$, the momentum cutoff, is of the order the
inverse of the domain wall thickness. The case $r\ll1$ where
dipolar interactions can be neglected has been already analyzed
in \cite{vazquez}, here we focuses our attention in the opposite
case $r\gg1$ where dipolar interactions are relevant. In this
case, imposing scale invariance we obtain that ${\cal N}$ and
$v$ should goes to zero and
\begin{equation}
z=1,\ \ \ \ \zeta=\frac{2-d}{2}.
\label{eq:10}
\end{equation}
The scaling exponents obtained in this way are identical to
those derived by CZDE \cite{cizeau} for the case of constant
magnetic field. Moreover the upper critical dimension $d_c=2$ is
also the same.

The saturation magnetization is not a control parameter because
the dipolar interaction term, which is proportional to $M_s^2$,
does not renormalize. Thus, ${\cal N}$ and $v$ are the only
control parameters. Since scale invariance is obtained when
${\cal N}=0^+$ from eq. (\ref{eq:9}) one can define the
correlation length $\xi\sim{\cal N}^{-\nu}$ with $\nu=1$.
Moreover, from eq. (\ref{eq:8}) we obtain that the
susceptibility scale as $\chi\sim{\cal N}^{-\gamma}$ with
$\gamma=1$. At the critical state $\xi\sim L$ and
\begin{equation}
\chi\sim\xi^{\gamma\nu}\sim L,
\label{eq:11}
\end{equation}
On the other hand, the increase of $v$ may
change the character of the noise correlator. For large $v$ the
noise correlator can be approximated by an annealed noise. In
this case the magnetic field predominates over disorder,
corresponding with a supercritical regime. From eq. (\ref{eq:9})
one can define the characteristic velocity
$v_c\sim\epsilon^\theta$ with
\begin{equation}
\theta=\nu(z-\zeta),
\label{eq:12}
\end{equation}
which divides the phase diagram $({\cal N},v)$ into two regions,
the supercritical state $v\gg v_c$ and the subcritical one $v\ll
v_c$. Criticality is obtained when ${\cal N}\rightarrow0$ and
$v\rightarrow0$.

To go further we have performed a RG analysis of the problem. We
integrate out the degrees of freedom in a momentum shell near
the cutoff $\Lambda$ and rescale $k\rightarrow b^{-1}k$,
$\omega\rightarrow b^{-z}\omega$, and $\tilde{w}\rightarrow
b^{\zeta+d+z}\tilde{w}$, where $b=\text{e}^l$ with
$l\rightarrow0$. The flow equations for the parameters $\Gamma$,
$\lambda$, ${\cal N}$, $M_s$ and $v$ are obtained through a
direct application of the RG transformations to eq. (\ref{eq:4})
in the Fourier space. The renormalization of the moments of the
noise correlator ($Q_{n}=\int_q\tilde{\Delta}(q)q^n$) is
obtained considering vertex functions \cite{review,fisher}. As
in perturbation theory here we only consider the case $r\gg1$.
In this case we found corrections of the order of $r^{-1}$ to
the exponents computed by perturbation theory \cite{vazquez1}.
Thus, the exponents $z$ and $\zeta$ in eq. (\ref{eq:10}) are
exact up to ${\cal O}(r^{-1})$.

Let us now determine the avalanche distribution exponents. The
Barkhausen signal $V(t)$ is the voltage produced from a pickup
coil around a ferromagnet subjected to a slowly varying applied
field. In the low-frequency limit the time scale for domain wall
motion is much smaller than the time between jumps and,
therefore, one may guarantee that each induced voltage jump
corresponds with a single avalanche in the domain wall motion. A
resolution voltage level $V_R$ is defined, such that one can not
resolve details below $V_R$. An elementary Barkhausen jump can
thus be defined as the portion of the $V(t)$ signal delimited by
two subsequent intersections of the signal with the $V_R$ line.
With this definition, the duration $T$ is simply the time
interval between these two subsequent intersections and the size
$s$ is the area delimited by $V(t)$ and $V_R$ between the same
points.

In the subcritical regime the dynamics takes place in the form
of avalanches, characterized by the avalanche size
$P(s)=s^{-\tau}f(s/s_c)$ and duration $P(T)=T^{-\alpha}g(T/T_c)$
distributions, where $s_c$ and $T_c$ are the avalanche size and
duration cutoffs. In the subcritical state
$s_c\sim\epsilon^{-1/\sigma}$ and $\xi\sim\epsilon^{-\nu}$ while
at criticality $s_c\sim L^D$ and $\xi\sim L$, where $D$ is the
avalanche dimension, leading to the scaling relation
\begin{equation}
\sigma=\frac{1}{D\nu}.  
\label{eq:14}
\end{equation}
Other scaling relations are obtained taking into account that
$\chi=\langle s\rangle$ and $\int dsP(s)=\int dTP(T)$, which
lead to
\begin{equation}
\gamma=\frac{(2-\tau}{\sigma},\ \ \ \ (\tau-1)D=(\alpha-1)z, 
\label{eq:15}
\end{equation}
respectively.

Using these scaling relations and our result $\gamma=\nu=z=1$ we
obtain the following scaling relations for the avalanche
exponents
\begin{equation}
\tau=2-\frac{1}{\alpha},\ \ \ \ 
\alpha=D,
\label{eq:16}
\end{equation}
On the other hand, for $d<d_c$, the avalanche dimension and the
roughness exponent are related via $D=d+\zeta$, while above the
upper critical dimension one obtains $D=d_c=2$. Hence, we can
compute $\tau$ and $\alpha$ using eq. (\ref{eq:16}) and this
value of $D$. For instance
\begin{eqnarray}
\alpha& =\frac{3}{2},\ \text{for}\ d+1=2;\nonumber\\
 & =2,\ \text{for}\ d+1\geq3.
\label{eq:17}
\end{eqnarray}

The scaling law in eq. (\ref{eq:11}), the scaling relations in
eq. (\ref{eq:16}) and the values of the scaling exponent
$\alpha$ in eq. (\ref{eq:17}) are also obtained for DASM
\cite{dhar}. However, the upper critical dimension of DASM is 3,
and not 2 as obtained here. This apparent contradiction can be
understood if one takes into account that in DASM time evolution
does not introduce an additional dimension because it can be
represented by the evolution in the preferential dimension. On
the contrary in the motion of the domain wall the time evolution
take place perpendicular to the $d$-dimensional substrate and,
therefore, introduces an additional dimension. Taking this fact
into account the upper critical dimension will be $2+1$, as in
DASM. Thus, the CZDE model in $d+1$ dimensions is
equivalent to a $d$-dimensional DASM.

The fact that the CZDE is mapped into a directed sandpile model,
and not into an undirected one, is due to dipolar interactions
which introduces an anisotropy in the system. If dipolar
interactions becomes negligible ($r\ll1$) then we expect that
the CZDE model will be mapped into an undirected sandpile model.
Experiments in magnetostrictive materials, where dipolar
interactions can be neglected, have been performed
\cite{urbach,bahiana,durin}. Earlier measurements by Urbach {\em
et al} \cite{urbach} gives $\tau=1.33\pm0.10$. More recently,
Durin and Zapperi (DZ) \cite{durin} reported the more accurate
exponents $\tau=1.28\pm0.02$ and $\alpha=1.5\pm0.1$. On the
other hand, numerical simulations of the   CZDE model in two
dimensions and without dipolar interactions have also been
performed \cite{bahiana,durin}. The more accurate numerical
estimates, reported by Durin and Zapperi \cite{durin}, are
$\tau=1.26\pm0.04$ and $\alpha=1.40\pm0.05$. These exponents are
in the range reported for undirected Abelian sandpile models in
two dimensions. For instance, numerical simulations of the
Bak-Tang-Wiesenfeld (the prototype of undirected Abelian
sandpile model) gives $\tau=1.293$ and $\alpha=1.480$
\cite{lubeck}. However, to validate this guess we cannot base
our analysis only in the avalanche exponents, we must provide
further elements. In \cite{vazquez} we have shown that the CZDE
without dipolar interactions can be mapped into a class of
undirected sandpile models with annealed noise in the toppling
rule.

In conclusion we have shown that the CZDE model for the dynamics
of a domain wall belongs to the universality class of directed
Abelian sandpile models. Dipolar interactions introduce an
anisotropy in the systems leading to a preferential direction
along the magnetization. We have provided strong arguments which
states the equivalence between the dynamics of a domain wall and
sandpile models. This conclusion can be extended to systems with
many domain walls, since for short time scales the interaction
between domain walls can be neglected and the single domain wall
picture is correct. Hence, the Barkhausen effect actually
exhibits self-organized critical behavior.

\end{multicols}


\begin{thebibliography}{50}

\bibitem{bak} P. Bak, C. Tang, and K. Wiesenfeld, Phys. Rev.
Lett. {\bf 59}, 381 (1987); Phys. Rev. A {\bf 38}, 364 (1988).

\bibitem{vespignani} A. Vespignani and S. Zapperi, Phys. Rev.
Lett. {\bf 78}, 4793 (1997); Phys. Rev. E {\bf 57}, 6345 (1998).

\bibitem{geoffroy} O. Geoffroy and J. L. Corteseil, J. Magn.
Magn. Mate. {\bf 97}, 198 (1991); {\em ibid} {\bf 97}, 205
(1991).

\bibitem{cote} P. J. Cote and L. V. Meisel, Phys. Rev. Lett.
{\bf 67}, 1334 (1991).

\bibitem{obrien} K. P. O'Brien and M. B. Weissman, Phys. Rev. E
{\bf 50}, 3446 (1994).

\bibitem{spasojevic} D. Spasojevi\'c, S. Bukni\'c, S.
Milo\v{s}evi\'c, and H. E. Stanley, Phys. Rev. E {\bf 54}, 2531
(1996).

\bibitem{sethna} J. P. Sethna, K. Dahmen, S. Kartha, J.
Krumahansl, B. W. Roberts, and J. D. Shore, Phys. Rev. Lett.
{\bf 21}, 3347 (1993).

\bibitem{perkovic} O. Percovi\'c, K. Dahmen, and J. P. Sethna,
Phys. Rev. Lett. {\bf 75}, 4528 (1995).

\bibitem{dahmen} K. Dahmen and J. Sethna, Phys. Rev. Lett. {\bf
71}, 3222 (1993); Phys. Rev. B {\bf 53}, 14872 (1996).

\bibitem{cizeau} P. Cizeau, S. Zapperi, G. Durin, and H. E.
Stanley, Phys. Rev. Lett. {\bf 79}, 4669 (1997); S. Zapperi, P.
Cizeau, G. Durin, and H. E. Stanley, Phys. Rev. B {\bf}, 1998.

\bibitem{alessandro} B. Alessandro, C. Beatrice, G. bertotti,
and A. Montorsi, J. Appl. Phys. {\bf 68}, 2901 (1990); {\em
ibid} {\bf 68}, 2908 (1990).

\bibitem{urbach} J. S. Urbach, R. C. Madison, and J. T. Markert,
Phys. Rev. Lett. {\bf 75}, 276 (1995).

\bibitem{note} When dipolar interactions are negligible a
different universality class is obtained.

\bibitem{dhar} For a review see D. Dhar, cond-mat/9808047.

\bibitem{fisher} O. Narayan and D. S. Fisher, Phys. Rev. Lett.
{\bf 68}, 3615 (1992); Phys. Rev. B {\bf 48}, 7030 (1993).

\bibitem{review} T. Nattermann, S. Stepanow, L.-H. Tang, and  H.
Leschhorn, J. Phys. II France {\bf 2},1483 (1992); H. Leschhorn,
T. Nattermann, S. Stepanow, and L.-H. Tang, Ann. Phys. {\bf 6},
1 (1997).

\bibitem{vazquez} A. V\'azquez and O. Sotolongo-Costa,
cond-mat/9812173.

\bibitem{vazquez1} A. V\'azquez and O. Sotolongo-Costa, in
progress.

\bibitem{bahiana} M. Bahiana, B. Koiller, S. L. A. de Queiroz,
J. C. Denardin, and R. L. Sommer, cond-mat/9808017.

\bibitem{durin} G. Durin and S. Zapperi, cond-mat/9808224.

\bibitem{lubeck} S. L\"ubeck and K. Usadel, Phys. Rev. E {\bf
55}, 4095 (1997).

\end{thebibliography}
\end{document}